\documentstyle[12pt,epsf]{article}
\def\beq{\begin{equation}}
\def\eeq{\end{equation}}
\def\beqa{\begin{eqnarray}}
\def\eeqa{\end{eqnarray}}
\def\a {{\rm f}}

\def\Qg{Q_{\rm gap}}

\begin{document}

\begin{flushright}
ITP-SB-99-57
\end{flushright}

\vbox{\vskip 0.1 true in}

\begin{center}
{{\Large \bf Interjet Rapidity Gaps in
Perturbative QCD}\footnote{Talk presented at {\it QCD and Multiparticle
Production}, XXIX International Symposium on Multiparticle
Dynamics, Brown University, Aug.\ 9-13, 1999.}}

\vbox{\vskip 0.1 true in}

{\large Gianluca Oderda and George Sterman}

\vbox{\vskip 0.2 true in}

{\small \it C.N.\ Yang Institute for Theoretical Physics \\
SUNY at Stony Brook,
Stony Brook, NY 11794-3840, USA}

 \end{center} 


\abstract{
We discuss a formalism in which high-$p_T$ dijet rapidity gaps
are identified by energy flow in the interjet
region.  When the gap energy, $Q_{\rm gap}$, is sufficiently
large, 
the cross section may be computed
from standard factorization theorems.  For
$p_T\gg Q_T\gg\Lambda_{\rm QCD}$,
this is a two-scale perturbative problem, in which we may
resum logarithms of $Q_{\rm gap}/p_T$.  
The cross section computed as a function of 
$Q_{\rm gap}$ reproduces many of the features of
the Tevatron dijet gap data.
The factorized cross section gives meaning to the color content of
the hard scattering.}

\section{Dijet Gaps and Color}

One of the enduring themes of high energy physics
 is scattering through the exchange of composite systems.
An exchange between incoming hadrons
that produces two sets of outgoing hadrons separated by a
large gap in rapidity is necessarily color singlet.  
Perturbation theory becomes relevant
when one end of the exchange involves a
hard scattering, producing jets or heavy quarks.  
In this talk, we discuss the
application of perturbative methods to exchanges 
that are hard at {\it both}  ends, and in which
high-$p_T$ jets are observed at large rapidity in
both the forward and backward directions, with
little or no particle multiplicity between.  These are
often called ``dijet rapidity gap" events.

As a heuristic principle, based in part on the antenna patterns of
QED, we expect the exchange of gluons in a
color singlet state to produce very little radiation
in the intermediate rapidity region \cite{B}. This
idea has such appeal that dijet gap events are 
routinely termed ``color singlet exchange".  To the extent that
it is not a truism, this term is
a description of the color of gluons exchanged at {\it short}
distances and times.  The idea of color singlet
exchange, however, has been a difficult one to implement in 
perturbative terms.
After all, gluons of any energy carry octet color charge, so that there
is no unique way of defining color exchange 
in a finite amount of time \cite{OZ}.
On the other hand, it takes a very short time to radiate
a hard gluon, and once radiated, it cannot
be reabsorbed on the basis of soft color rearrangements
at very long times.   

\section{The Two-Gluon Model, Soft Color and Factorization}

The simplest short-distance model for dijet gaps is 
based on two-gluon exchange \cite{B}. In  a two-gluon model, the
gap is usually filled by spectator interactions, up to a
``survival probability",  $P_S$, which may be estimated \cite{B,GLM}
from low-$p_T$ diffractive scattering to be of order 0.1.
Denoting the probability for
hard color-singlet exchange as $f_1$, the fraction
of gap events becomes
\beq
f_{\rm gap}=f_1\; P_S\, .
\label{gapfrac}
\eeq
If we estimate $f_1\sim {\cal O}(\alpha_s(p_T)/\pi)\sim 0.1$, we predict gap events
at the one percent level, and this is what is 
seen experimentally \cite{CDF,D0,630,ZEUS}.
This analysis would lead us to expect more gap events from
gluon-gluon than quark-quark scattering, because of the larger
color factors in two-gluon exchange graphs for 
the former.  This expectation was tested by comparing
 630 and 1800 GeV data from
the Tevatron,  because at fixed $p_T$ the role of gluon-gluon scattering 
increases with the overall center-of-mass energy.  
The proportion of gap events, however, decreased, rather than increased,
with the energy.

In the soft color approach \cite{EGH}, normally presented as an 
alternative to the two-gluon model, the underlying
hard scattering is treated at lowest order, which
for gap events is primarily single-gluon, color octet exchange.
The gap probability is determined by counting possible
color exchanges, assuming all to be equally likely, up to
an overall survival factor (rather larger than 1/10).
Because gluons have more color states than quarks, they
are correspondingly less likely to produce gap events.
The soft color  model then naturally leads to fewer gap events
as the energy, and hence the role of gluons, increases.

Clearly, dijet gaps provide insight into a tantalizing mixture of
short- and long-distance physics.  The successes of the soft color
picture suggest, however, that to learn more about color
exchange at short distances, we must generalize the two-gluon 
model.  From the point of view of perturbative QCD this is
already necessary, because two-gluon exchange is not, by itself,
infrared safe.  That is, even at lowest order it is sensitive
to long-distance configurations where one gluon carries almost
all the momentum transfer, and the other almost none.
It is therefore natural  to
seek a definition of color exchange 
that is infrared
safe, and hence well-defined in perturbation theory.   
To do so, we find it useful to extend the concept
of gap events.

We have developed a (resummed) perturbative
QCD formalism for dijet rapidity gaps, made possible by
redefining gaps in terms of an energy flow, $Q_{\rm gap}$, rather than 
particle multiplicty \cite{SO}. 
The resulting cross sections can be treated via standard factorization
theorems. In this formulation, if $\Qg\gg \Lambda_{\rm QCD}$ 
the cross section is perturbatively calculable.  In addition,
when $p_T\gg \Qg\gg \Lambda_{\rm QCD}$, our gap cross sections have
two perturbative scales, and logarithms in their ratio can be
resummed by renormalization group methods.  

Resummation in $\ln(\Qg/p_T)$
allows us to probe color flow at short distances, and to 
generalize the concept of hard color singlet exchange.
As we shall see, for Tevatron kinematics, the dominant short-distance color
exchange in rapidity gap events has a healthy proportion of color octet.

The dijet cross section at measured $\Qg\gg\Lambda_{\rm QCD}$ falls into the
class of inclusive jet cross sections that can be written in
factorized form:
\beqa
\frac{d\sigma}{d\Qg\, d\cos\hat{\theta}}\left(S,E_{T},\Delta y\right) 
&=&
 \sum_{f_A,f_B} \int d\cos\hat{\theta}\; \int_0^1 dx_A \int_0^1 dx_B\;
 \phi_{f_A/p}(x_A,-\hat{t}) \,
\phi_{f_B/\bar{p}}(x_B,-\hat{t}) \nonumber \\
&& \hspace {10mm} \times \sum_{f_C,f_D}\frac{d\hat{\sigma}^{(\a)}}
{d\Qg \, d\cos\hat{\theta}} 
\left(\hat{t},\hat{s},y_{JJ},\Delta y,\alpha_s(\hat{t})\right)\, ,
\label{crosssec}
\eeqa
with $\phi_{f/h}$ parton distributions, evaluated at scale $\sqrt{-\hat t}$,
the dijet momentum transfer. The partonic
cross section ${d\hat{\sigma}^{(\a)}}/{d\Qg \, 
d\cos\hat{\theta}}$
is a hard scattering function, starting with the Born cross section
at lowest order (cf. the soft color model).  The 
index $\a$ denotes the partonic hard scattering
$f_A+f_B \rightarrow f_C +f_D$. 
The cross section depends on the dijet pair rapidity $y_{JJ}$, 
the partonic center-of-mass (c.m.) energy squared $\hat{s}$, 
the partonic c.m. scattering angle $\hat{\theta}$, with
$-\frac{\hat{s}}{2}\left(1-\cos \hat{\theta} \right)=\hat{t}$,
and $\Delta y$, the gap size as a rapidity interval.

\section{Refactorization and Resummation}

Our central observation is that we may further ``refactorize"
the hard-scattering cross section in Eq. (\ref{crosssec})
to separate the gap energy dependence, at scale $\Qg\gg\Lambda_{\rm QCD}$, from the underlying
hard scattering, at scale $p_T\gg \Qg$.   This factorization depends
upon the color exchange in the residual hard scattering.  Taking into
account the hard-scattering amplitude and its complex conjugate, we
have the following matrix relation for $\hat \sigma$ in (\ref{crosssec}):
\beqa
\Qg\frac{d\hat{\sigma}^{(\a)}}{d\Qg\, d\cos\hat{\theta}}\left(\hat{s},\hat{t},y_{JJ},\Delta y,\alpha_s(-\hat t)\right)
&=&
H_{IL}\left( \frac{\sqrt{-\hat{t}}}{\mu},
\sqrt{\hat{s}},\sqrt{-\hat t},\alpha_s(\mu^2) \right) \nonumber \\
&&  \times S_{LI} \left( \frac{\Qg}{\mu},y_{JJ},\Delta y\right)\, ,
\label{factor}
\eeqa
where $H_{IL}$ incorporates scattering at scale $p_T$, and $S_{LI}$ 
the intermediate radiation scale $\Qg$.  Because
our cross section is inclusive, dependence on momentum scales below $\Qg$
cancels.  The parameter $\mu$ is a new ``refactorization scale",
which must be introduced to  define the tensors $H_{IL}$ and $S_{LI}$.
Their indices refer to a basis of color exchanges
in the amplitude and the complex conjugate at the shortest distance
scale (interference is possible).  For example, in quark-antiquark scattering, we
may choose the basis: $I,J=$ $t$-channel singlet, $t$-channel octet.

The left-hand side of the refactorization relation Eq.~(\ref{factor}) is
independent of the new factorization scale $\mu$, a requirement that
immediately leads to renormalization group equations, \cite{color}
\beq 
\left(\mu\frac{\partial}{\partial\mu}+\beta(g)\frac{\partial}{{\partial}g}
\right)S_{LI}=
-(\Gamma_S^\dagger)_{LB}S_{BI}-S_{LA}(\Gamma_S)_{AI}\, ,
\label{eq:resoft}
\eeq
for $S$, and similarly for $H$, where $\Gamma_S(\alpha_s)$ is an anomalous dimension
matrix that has been computed at one loop \cite{SO,O}.  
Thus, the color content  of  the hard scattering $H_{IL}$
determines the $\Qg$-dependence of the cross section, through 
the eigenvalues of $\Gamma_S$, which themselves depend  on the
flavors and scattering angles of the underlying hard scattering.
To leading logarithm in $\Qg/\sqrt{-\hat{t}}$, we find
\beqa
\frac{d\hat{\sigma}^{(\a)}}{d\Qg \, d\cos \hat{\theta}}\left(\hat{s},\hat{t},y_{JJ},\Delta y,\alpha_s(-\hat t)\right)
&=& 
\nonumber \\
&& \hspace{-40mm} H^{(1)}_{\beta \gamma}\left( \Delta y,\sqrt{\hat{s}},\sqrt{-\hat{t}},
\alpha_s\left(-\hat{t}\right) \right) 
 S^{{(0)}}_{\gamma \beta} ( \Delta y ) \,  \nonumber \\
&& \hspace{-35mm} \times {E_{\gamma\beta} \over \Qg}\; 
\left[\ln\left({\Qg\over \Lambda}\right)\right]^{E_{\gamma\beta}-1}\;  
\left[ \ln \left( {\sqrt{-\hat{t}}\over \Lambda}\right)\right]^{-E_{\gamma\beta}}\, . 
\label{factor2}
\eeqa
In this expression, written in the color tensor basis that diagonalizes
the anomalous dimension matrix $\Gamma$, the exponents $E_{\gamma \beta}$ are given
in terms of the eigenvalues,
$\lambda_\beta=(\alpha_s/\pi)\lambda^{(1)}_\beta$, of $\Gamma$ by
\beqa
&&E_{\gamma\beta}\left(y_{JJ},\hat{\theta},\Delta y \right)
=\frac{2}{\beta_0}\, \left[{ \lambda}^{(1)}{}^{*}_{\gamma} \left(y_{JJ},
\hat{\theta},
\Delta y \right) +\hat{ \lambda}^{(1)}_{\beta} \left( y_{JJ},\hat{\theta}.
\Delta y \right) \right]\, ,
\label{expon}
\eeqa
with $\beta_0=11N_c/3-2n_f/3$.
The linear combination of color exchanges with smallest eigenvalue
thus dominates the behavior of the cross section 
in the limit $\Qg/p_T\to 0$.

The concept of a dominant  eigenvalue generalizes conventional hard singlet
exchange, because the eigenvectors of the anomalous dimension matrix are
linear combinations of elements in the basis of $t$-channel color 
transfers.  The coefficients depend, in general, on the scattering angle
of the hard scattering.  Eq. (\ref{expon}) thus leads to a 
detailed set of predictions for dijet data with measured interjet 
energy flow.  

\section{Rapidity and Gap Energy Dependence}

Explicit forms of the anomalous dimension eigenvalues $\lambda_\alpha$
for quark and gluon processes 
may be found in Ref. \cite{O}.  Here, we would like only to
emphasize a few general features.  First, 
the overlap of the dominant eigenvector with hard color singlet exchange
grows in the direction of forward scattering, so that in the Regge
limit $-\hat{t}/\hat{s}\to 0$, $\hat{t}$ fixed, the dominant color
exchange becomes purely color singlet \cite{DDT}. Second, the eigenvalues
for gluon-gluon scattering are larger than those for 
processes involving quarks.  This
makes it harder for gluon-gluon hard scattering to produce 
rapidity gaps, for much the same reasons as in the soft color model.
The size of the eigenvalue $\lambda_\alpha$ is related to the
number of color states available.

A typical prediction is shown in Fig. 1 \cite{SO}, where we illustrate
the differential cross section for measured $\Qg$ at 630 and 1800 GeV, 
the former with a gap of 3.2 units of rapidity, and latter at
4 units, following the D0 experiment.  These curves were  generated
from valence quarks and antiquarks only, so at any scattering angle
color exchange is a linear combination of singlet 
and octet.  For any of the scattering angles allowed 
by the kinematics, the smallest (``dominant") eigenvalue 
as $\Qg\to 0$ is primarily, but not completely, singlet, and the
other primarily octet.  We refer to them ``quasi-singlet" and ``quasi-octet",
respectively.
The projection of the hard scattering on the $t$-channel eigenvectors is nonvanishing
for both quasi-singlet and quasi-octet, but is larger for quasi-octet.  
Examining Eq.~(\ref{expon}), we see that the $\Qg$-dependence of 
the cross section for each color exchange is determined by the
size of the eigenvalues, along with the projection.  

The quasi-singlet contribution, shown by dashed lines in
the figure, has
an eigenvalue that is less than unity in absolute value.
For each value of $-\hat{t}$ it is therefore a monotonically
decreasing function of $\Qg$.   The differential cross section has  a weak
singularity at $\Qg=\Lambda_{\rm  QCD}$, but its moments are
calculable.  Quasi-octet exchange (dotted lines) on the other  hand, has an eigenvalue
that  is always greater than unity.  For small $\Qg>\Lambda_{\rm QCD}$, it therefore
increases in $\Qg$ until the dimensional factor $1/Q_c$ takes over.
The sum of these two contributions, along with their nonvanishing interference,
is shown by the solid lines in the figure.  For most of the range
in $\Qg$, it has the same shape as the quasi-octet curve:
small for small $\Qg$, reaching a  peak at order $\sqrt{-\hat{t}}$,
and then decreasing.  At low gap energy, however, the 
quasi-singlet exchange produces a 
small upturn.  This is the ``hard singlet exchange"
observed by CDF and D0 \cite{CDF,D0}.  Gap events defined by vanishing particle
multiplicity in the interjet region are counted in this excess.
Our prediction for such events must be  diluted, as usual, by
corrections associated with spectator interactions, which, according
to the factorization formalism, are suppressed only by powers
of $\Lambda_{\rm QCD}/\Qg$, and which therefore become important
for small $\Qg$.
We have in Eq.~(\ref{crosssec}), however, a set of predictions for the
full range of $\Qg$.

Recently, gluonic processes were considered as part of a complete analysis of dijet
gaps in photoproduction, seen by Zeus \cite{O}.  As noted above, for
gluon-gluon scattering the eigenvalues are
larger than for quark-antiquark scattering, and, indeed, are
greater than  unity over the accessible phase space.  
As a result, gluonic processes lack the upturn we have just
found at low $Q_{\rm gap}$ for quark processes.  This is the
explanation for the lower gap cross section observed as the
role of gluons increases \cite{630}.

\begin{figure}[t]
\begin{center}      
\mbox{\epsfysize=5.5cm \epsfxsize=11cm \epsfbox{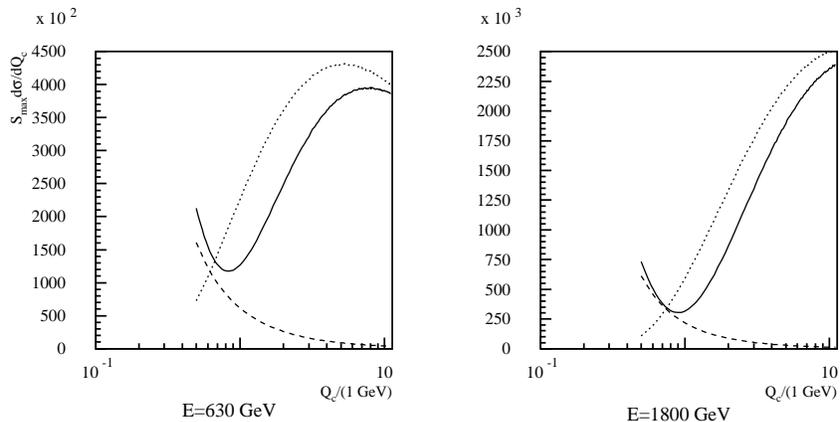}}
\end{center}
\caption{The cross section as a function of gap energy, labelled $Q_c$
(solid line) and contributions from
``quasi-octet" (dotted line) and quasi-singlet (dashed line)
color exchange, for
$630$ and $1800$ GeV. Units are arbitrary.}
\label{cteqSinglejet}
\end{figure}

\section{Summary}

Energy flow analysis makes possible a quantitiative
study of radiation in interjet regions, and gives a perturbative
meaning to short-distance color exchange, generalizing
the two-gluon exchange model.  On the basis of this
analysis, gaps in dijet events come from a compound structure,
predominantly,  but not purely, singlet in the hard scattering.
Many qualitative results, including the relative suppression of dijet
gaps for gluon-gluon scattering, are similar to those of the
soft color model.  Although much more remains to be done,
the perturbative analysis offers a systematic set
of differential predictions for energy flow, as
a function of momentum transfer, flavor and gap width.
In principle,  these ideas can  be tested at Run II of the
Tevatron, at  Hera and at the LHC.

\subsection*{Acknowledgments}
We thank Chung-I Tan and Alan White for their invitation to
present this work at the ISMD conference.
 This work was supported in part by the National Science Foundation, grant PHY9722101.

\end{document}